\documentclass[reprent,aps,prb,onecolumn,superscriptaddress]{revtex4-2}

\usepackage{amsmath,amssymb}
\usepackage{graphicx}
\usepackage{bm}
\usepackage{multirow}
\usepackage{hyperref}
\usepackage{wasysym}
\usepackage{graphicx}
\usepackage{dcolumn}
\usepackage{physics}

\usepackage{color}

\newcommand{\Eq}[1]{Eq.~(\ref{#1})}

\newcommand{\Fig}[1]{Fig.~\ref{#1}}

\newcommand{\Figure}[1]{Figure~\ref{#1}}

\usepackage{orcidlink}

\begin{document}

\preprint{APS/123-QED}

\title{Square skyrmion lattice in multi-orbital $f$-electron systems}

\author{Yan Zha\, \orcidlink{0009-0008-8568-446X}}
\email{yzha@phys.sci.hokudai.ac.jp}
\author{Satoru Hayami\, \orcidlink{0000-0001-9186-6958}}
\email{hayami@phys.sci.hokudai.ac.jp}
\affiliation{Graduate School of Science, Hokkaido University, Sapporo 060-0810, Japan}

\date{\today}
\begin{abstract}
\footnote{This version is content-wise identical to the published version \href{https://journals.aps.org/prb/abstract/10.1103/PhysRevB.111.165155}{[Yan Zha and Satoru Hayami, Phys. Rev. B {\bf 111}, 165155 (2025)]}.}
We report the emergence of a square-shaped skyrmion lattice in multi-orbital $f$-electron systems with easy-axis magnetic anisotropy on a centrosymmetric square lattice.
By performing mean-field calculations for an effective localized model consisting of two Kramers doublets, we construct the low-temperature phase diagram in a static external magnetic field.
Consequently, we find that a square-shaped skyrmion lattice with the skyrmion number of one appears in the intermediate-field region when the crystal field splitting between the two doublets is small.
Furthermore, we identify another 
double-$Q$ state with
a nonzero net scalar chirality at
zero- 
and low-field regions,
which is attributed to the help of the multi-orbital degree of freedom.
Our results offer another route to search for skyrmion-hosting materials in centrosymmetric $f$-electron tetragonal systems with multi-orbital degrees of freedom, e.g., Ce-based compounds.
This contrasts with conventional other $f$-electron systems hosting skyrmion lattices, such as Gd- and Eu-based compounds without orbital angular momentum.
\end{abstract}
\maketitle

\section{INTRODUCTION} \label{sec:introduction}
Skyrmions, a class of quasi-particles, were first proposed by Tony Skyrme in the 1960s~\cite{SKYRME1962556}, describing a type of topologically stable field configuration.
Subsequently, they have become a significant topic in condensed matter physics~\cite{Bogdanov89, BOGDANOV1994255, roessler2006spontaneous,nagaosa2013topological,zhang2020skyrmion,gobel2021beyond,hayami2021topological}.
Due to their topological nature, skyrmions remain invariant under continuous deformations~\cite{nagaosa2013topological}, which has led to active research into their stability
~\cite{zhang2016antiferromagnetic, oike2016interplay,cortes2017thermal,je2020direct}.
Their topological robustness and unconventional transport properties have garnered increasing attention in spintronics, and skyrmions hold great promise for applications, making them strong candidates for next-generation computing and storage devices~\cite{fert2013skyrmions,romming2013writing,fert2017magnetic,zhang2020skyrmion}.

Compared to a single skyrmion, a skyrmion lattice (SkL) is a highly ordered structure as a thermodynamic phase.
The conventional microscopic mechanism for the formation of SkLs is the synergy of the ferromagnetic exchange interaction, the Dzyaloshinskii--Moriya (DM) interaction~\cite{DZYALOSHINSKY1958241, PhysRev.120.91}, and the Zeeman coupling induced by a static external magnetic field in noncentrosymmetric magnets
~\cite{Bogdanov89, BOGDANOV1994255,roessler2006spontaneous, Yi_PhysRevB.80.054416}.
This naturally prompts the question of whether SkLs can arise in a system without spatial inversion symmetry breaking.
Experimental advances in recent years have shown that centrosymmetric materials with spatial inversion symmetry can indeed host SkLs under an external magnetic field.
Examples include Gd$_2$PdSi$_3$~\cite{kurumaji2019skyrmion,Hirschberger_PhysRevLett.125.076602,Hirschberger_PhysRevB.101.220401,Spachmann_PhysRevB.103.184424, Paddison_PhysRevLett.129.137202} on a triangular-lattice hexagonal structure
and Gd$_3$Ru$_4$Al$_{12}$~\cite{hirschberger2019skyrmion, Hirschberger_10.1088/1367-2630/abdef9, Nakamura_PhysRevB.107.014422} on a kagome-type hexagonal structure
and GdRu$_2$Si$_2$~\cite{khanh2020nanometric,khanh2022zoology, Wood_PhysRevB.107.L180402}, EuAl$_4$~\cite{Shang_PhysRevB.103.L020405,kaneko2021charge,takagi2022square,PhysRevB.106.094421,PhysRevB.107.L020410},
and GdRu$_2$Ge$_2$~\cite{yoshimochi2024multistep} on a centrosymmetric tetragonal lattice.
Although the DM interaction leading to the SkLs does not exist in systems with inversion symmetry, several microscopic mechanisms have been theoretically proposed to stabilize the SkLs~\cite{hayami2024stabilization},
such as competing exchange interactions~\cite{PhysRevLett.108.017206, leonov2015multiply, Lin_PhysRevB.93.064430},
dipolar interactions~\cite{UtesovPhysRevB.103.064414, UtesovPhysRevB.105.054435}, nonmagnetic impurity~\cite{Hayami_PhysRevB.94.174420}, crystal-dependent magnetic anisotropy~\cite{amoroso2020spontaneous, Hayami_PhysRevB.103.054422, Yambe_PhysRevB.106.174437, Zhang_PhysRevLett.133.196702}, and electric-field-induced three-spin interactions~\cite{yambe2024dynamical}. 

Specifically, we focus on the square-shaped SkL (S-SkL), where skyrmions are packed so as to satisfy the fourfold rotational symmetry on a two-dimensional square lattice. 
The S-SkL is described by a double-$Q$ state, which is formed by a superposition of two spiral waves with mutually perpendicular wave vectors.
From an energetic viewpoint, the formation of such a double-$Q$ S-SkL is severe compared to that of a triple-$Q$  triangular-shaped SkL with the ordering wave vectors $\bm{Q}_1$, $\bm{Q}_2$, and $\bm{Q}_3$, the latter of which has an effective fourth-order coupling in the free energy owing to the relation of $\bm{Q}_1+\bm{Q}_2+\bm{Q}_3=\bm{0}$~\cite{Lin_PhysRevB.93.064430, PhysRevB.93.184413}. 
Indeed, it has been revealed that the S-SkL in centrosymmetric systems appears in the ground state by considering additional effects, such as biquadratic interaction~\cite{Hayami_PhysRevB.103.024439}, compass-type anisotropic interaction~\cite{UtesovPhysRevB.103.064414, wang_PhysRevB.103.104408}, higher-harmonic wave vector interaction~\cite{hayami_PhysRevB.105.174437}, and
long-range magnetic anisotropy~\cite{Okigami_PhysRevB.110.L220405}. 
These theoretical studies offer microscopic mechanisms of the S-SkLs observed in experimental materials, as mentioned above, as well as candidate materials hosting the S-SkL, such as EuGa$_4$~\cite{Zhu_PhysRevB.105.014423,zhang2022giant}, EuGa$_2$Al$_2$~\cite{Moya_PhysRevMaterials.6.074201}, Mn$_{2-x}$Zn$_x$Sb~\cite{Nabi_PhysRevB.104.174419}, and GdOs$_2$Si$_2$~\cite{hayashi2024exploring}.

Meanwhile, one notices that, to date, most materials hosting the S-SkLs contain $4f$ lanthanoid elements without the orbital angular momentum like Gd and Eu ions.
This fact motivates us to explore whether the S-SkLs are possible in other $4f$-electron compounds with the orbital angular momentum, such as Ce ions.
Moreover, most previous studies have been performed 
for the effective spin models by renormalizing or ignoring the orbital degree of freedom. 
In other words, the multi-orbital effect on the S-SkLs has not been fully elucidated.

In the present study, we theoretically incorporate the multi-orbital effect in order to further understand the stabilization mechanism of the S-SkL in centrosymmetric hosts. 
We specifically consider the localized model consisting of two Kramers doublets with the $f^1$ configuration under strong easy-axis magnetic anisotropy as a consequence of the interplay between the spin--orbit interaction and crystalline electric field on the square lattice.
Within the mean-field calculations for the localized model, we show that several double-$Q$ states including 
the S-SkL and 
another 
double-$Q$ state with a net scalar chirality
emerge in the low-temperature phase diagram depending on the external magnetic field and crystalline electric field: The former is the fourfold-symmetric S-SkL with the skyrmion number of one, which is stabilized in the intermediate-field region and the
latter is the fourfold-asymmetric 
double-$Q$ state with inequivalent intensities at two ordering wave vectors,
which is stabilized from zero- to low-field region.
We identify these 
topologically nontrivial states
by examining the
structure factor, local and 
net scalar
chirality,
magnetization,
and topological skyrmion number.
Our results can be applicable to $4f$-electron compounds with the $f^1$ configuration like the Ce-based compounds.

The rest of the paper is organized as follows.
In Sec.~\ref{sec:model}, we introduce an effective
localized model that includes the exchange interaction, the Zeeman coupling, and the 
crystal field splitting
between the two Kramers 
doublets under the spin--orbit coupling and tetragonal crystalline electric field.
In Sec.~\ref{sec:method}, we present the numerical mean-field method used to investigate the ground state under different external magnetic fields and
crystal field splittings, and we provide multiple
physical quantities to collectively characterize 
the S-SkL
and the other 
multi-$Q$
magnetic-moment
configurations.
In Sec.~\ref{sec:results}, we report
the low-temperature phase diagram for the localized model, elucidate the mechanism behind the formation of the S-SkL, and discuss in detail the other magnetic phases.
Finally, in Sec.~\ref{sec:discussion and summary}, we summarize
the results of the present paper.
In Appendix~\ref{app1}, we show the derivation of the low-energy atomic bases in the effective localized model.
In Appendix~\ref{app2}, we
examine the
stability of the S-SkL phase when one of the crystal field parameters $\alpha$ is varied.

\section{MODEL} \label{sec:model}
We consider the situation where the $4f$ electrons are well localized at each lattice site on the two-dimensional square lattice. 
In addition, we suppose the $f^1$ configuration with the Ce$^{3+}$ ion in mind. 
When both effects of the atomic spin--orbit coupling and the tetragonal crystalline electric field are taken into account, the fourteen-degenerated energy levels with the total orbital angular momenta $J=7/2$ and $J=5/2$
are split into seven Kramers doublets, as detailed in Appendix~\ref{app1}. 
We construct an effective localized model by choosing two out of seven Kramers doublets, whose atomic bases are represented by using the notation $\ket{J, J_{z}}$ as 
\begin{align}
\label{eq:2KramersDoublets}
    \begin{cases}
        \ket{\Gamma_{t7 \pm}^{(1)}} 
        = \alpha \ket{\tfrac{5}{2},\pm \tfrac{5}{2}} 
          - \beta \ket{\tfrac{5}{2},\mp \tfrac{3}{2}}, 
        \\[10pt]
        \ket{\Gamma_{t7 \pm}^{(2)}} 
        = \beta \ket{\tfrac{5}{2},\pm \tfrac{5}{2}} 
          + \alpha \ket{\tfrac{5}{2},\mp \tfrac{3}{2}},
        \\[10pt]
    \end{cases}
\end{align}
where
\begin{align}
\label{eq:alpha}
    \alpha &= \frac{ 
    2\sqrt{5} \, B_4^4 }{\sqrt{ 
        \left( \sqrt{X^2 + 20\left(B_4^4\right)^2} + X \right)^{2} + 20\left(B_4^4\right)^2}}, \\
      X &= \left(B_{2}^{0} + 20\,B_{4}^{0}\right),\\
    \beta^2 &= 1 -\alpha^2, 
\end{align}
with the crystal field parameters $B_2^0$, $B_4^0$, and $B_4^4$ under the tetragonal symmetry (see Appendix~\ref{app1} in detail). 
Then, the effective localized Hamiltonian is given by
\begin{eqnarray} \label{eq:hamiltonian}
    \mathcal{H}_{\text{tot}} &=& \mathcal{H}_{\text{ex}} + \mathcal{H}_{\Delta} +  \mathcal{H}_{\rm Z},
    \\
    \mathcal{H}_{\text{ex}} &=& -\sum_{\left \langle i,j  \right \rangle }^{} \mathcal{J}_{ij}  \bm{J}_i \cdot \bm{J}_j , 
    \\
    \mathcal{H}_{\Delta} &=& 
        \left ( \begin{array}{cccc}
            0 & 0 & 0 & 0\\
            0 & 0 & 0 & 0\\
            0 & 0 & \Delta & 0 \\
            0 & 0 & 0 & \Delta
        \end{array} \right) , \label{eq:H_Delta} \\
        \mathcal{H}_{Z} &=& -h \sum_{i}^{}  J_{i}^{z},  
\end{eqnarray}
where the total Hamiltonian $\mathcal{H}_{\text{tot}}$ includes the contributions from the exchange interactions in $\mathcal{H}_{\text{ex}}$, crystal field splitting in $\mathcal{H}_{\Delta}$, and the Zeeman coupling in $\mathcal{H}_{\text{Z}} $. 
We suppose that the ground-state energy level is $ \ket{\Gamma_{t7\pm}^{(1)}}$ and the first-excited energy level is $\ket{\Gamma_{t7\pm}^{(2)}}$, as often found in Ce-based compounds, such as CeRhIn$_5$ and CeCoIn$_5$~\cite{PhysRevB.99.235143}.
Here, $\bm{J}_{i}$ represents the localized
total angular momentum operator $\bm{J}_{i}$ at site $i$, whose matrix elements for the bases $\left\{ \ket{\Gamma_{t7 +}^{(1)}}, \ket{\Gamma_{t7 -}^{(1)}},\ket{\Gamma_{t7 +}^{(2)}}, \ket{\Gamma_{t7 -}^{(2)}} \right\}$ are expressed as
\begin{widetext}
\label{eq:gt71gt72jxjyjz}
\begin{align}
    J^{x}= \left(
\begin{array}{cccc}
 0 & -\sqrt{5} \alpha  \beta  & 0 & \frac{1}{2} \sqrt{5} (\alpha -\beta ) (\alpha +\beta ) \\
 -\sqrt{5} \alpha  \beta  & 0 & \frac{1}{2} \sqrt{5} (\alpha -\beta ) (\alpha +\beta ) & 0 \\
 0 & \frac{1}{2} \sqrt{5} (\alpha -\beta ) (\alpha +\beta ) & 0 & \sqrt{5} \alpha  \beta  \\
 \frac{1}{2} \sqrt{5} (\alpha -\beta ) (\alpha +\beta ) & 0 & \sqrt{5} \alpha  \beta  & 0 \\
\end{array}
\right),
\end{align}

\begin{align}
     J^{y}=\left(
\begin{array}{cccc}
 0 & i \sqrt{5} \alpha  \beta  & 0 & -\frac{1}{2} i \sqrt{5} (\alpha -\beta ) (\alpha +\beta ) \\
 -i \sqrt{5} \alpha  \beta  & 0 & \frac{1}{2} i \sqrt{5} (\alpha -\beta ) (\alpha +\beta ) & 0 \\
 0 & -\frac{1}{2} i \sqrt{5} (\alpha -\beta ) (\alpha +\beta ) & 0 & -i \sqrt{5} \alpha  \beta  \\
 \frac{1}{2} i \sqrt{5} (\alpha -\beta ) (\alpha +\beta ) & 0 & i \sqrt{5} \alpha  \beta  & 0 \\
\end{array}
\right),
\end{align}

\begin{align}
    J^{z}=\left(
\begin{array}{cccc}
 \frac{1}{2} \left(5 \alpha ^2-3 \beta ^2\right) & 0 & 4 \alpha  \beta  & 0 \\
 0 & \frac{1}{2} \left(3 \beta ^2-5 \alpha ^2\right) & 0 & -4 \alpha  \beta  \\
 4 \alpha  \beta  & 0 & \frac{1}{2} \left(5 \beta ^2-3 \alpha ^2\right) & 0 \\
 0 & -4 \alpha  \beta  & 0 & \frac{1}{2} \left(3 \alpha ^2-5 \beta ^2\right) \\
\end{array}
\right).
\end{align}
\end{widetext}
It is noted that $\bm{J}$ is anisotropic in this Hilbert space, which arises from the spin--orbit coupling and crystalline electric field.

For the above atomic bases, we consider the exchange interactions between 
different sites, as represented by $\mathcal{H}_{\rm ex}$~\cite{PhysRevB.91.174438_Iwahara}.
In order to consider the finite-$q$ magnetic instability, we consider the frustrated exchange interactions consisting of the first-, second-, and third-neighbor interactions on the square lattice, whose coupling constants are denoted as $J_1$, $J_2$, and $J_3$, respectively. 
We choose these exchange parameters so that the ordering wave vectors are located at finite-$q$ positions in momentum space.
This is demonstrated by the
Fourier transformation of $\mathcal{H}_{\rm ex}$
leading to
\begin{align}\label{eq:h_ex_in_fourier}
   \mathcal{H}_{\text{ex}} 
   = - \sum_{\bm{q}} 
       \bm{J}_{\bm{q}} \cdot \bm{J}_{-\bm{q}} \,\mathcal{J}(\bm{q}),
\end{align}
where $\bm{q}$ represents the wave vector and
\begin{align}\label{eq:vecs_in_fourier}
    \bm{J}_{i} 
    = \frac{1}{\sqrt{N}} \sum_{\bm{q}} \bm{J}_{\bm{q}}
      \exp \left (i \,\bm{q}\cdot \bm{r}_{i}\right).
\end{align}
Here, $\bm{r}_i$ is the position vector of site $i$. 
From the above expression, the relation $\bm{J}_{\bm{q}}^{*} = \bm{J}_{-\bm{q}}$ holds, which ensures that $\bm{J}_{\bm{q}} \cdot \bm{J}_{-\bm{q}}$ is a real number.
The quantity $\mathcal{J}(\bm{q})$ is also real and defined as
\begin{align}\label{eq:Jq}
    \mathcal{J}(\bm{q}) 
    &= \sum_{\bm{r}_{i}-\bm{r}_{j}} 
        \mathcal{J}_{ij}(\bm{r}_{i}-\bm{r}_{j}) 
        \exp \left[ i\, \bm{q} \cdot (\bm{r}_{i}-\bm{r}_{j}) \right] 
        \nonumber \\
    &= \mathcal{J}(-\bm{q}) 
     = \mathcal{J}(\bm{q})^{*}.
\end{align}

For the square lattice with the interactions $J_1$--$J_3$, $\mathcal{J}(\bm{q})$ is explicitly expressed as
\begin{align}\label{eq:Jq_square}
\mathcal{J}(\bm{q}) &= 2J_1 (c_x + c_y) + 4J_2 c_{x}c_{y} + 2J_3 (c_{2x} + c_{2y}),
\end{align}
where $c_{\mu} = \cos(a\,q_{\mu})$ and $c_{2\mu} = \cos(2a\,q_{\mu})$ for $\mu = x, y$;
$a$ denotes the lattice constant of the square lattice, and we set $a = 1$ for simplicity.  
When the second and third terms in Eq.~(\ref{eq:hamiltonian}) are absent, the ordering wave vectors of the ground-state magnetic states are determined by the maximum value of $\mathcal{J}(\bm{q})$.
We choose
the short-period ordering wave vectors
as $\bm{Q}_1 = \left(\frac{2\pi}{6}, \frac{2\pi}{6}\right)$
so as to keep computational costs relatively low, where $J_1$ is the energy unit of the model; the ground state is characterized by a helical state with $\bm{Q}_1$ or $\bm{Q}_2=\left(-\frac{2\pi}{6}, \frac{2\pi}{6}\right)$ when the interaction is isotropic~\cite{doi:10.1143/JPSJ.14.807}. 
Although there are several combinations of ($J_1, J_2, J_3$) to give the ordering wave vector at $\bm{Q}_1$, we choose $J_1=1$, $J_2=-0.5$, and $J_3=-0.25$ so that the energy contribution from the $\bm{Q}_1+\bm{Q}_2$ becomes large, since such a contribution plays an important role in enhancing the instability toward the
SkL compared to the helical state~\cite{hayami_PhysRevB.105.174437}.

The effect of crystal field splitting between two Kramers doublets is expressed as $\mathcal{H}_{\Delta}$. 
Although the parameter $\Delta$ is related to the crystal field parameters, $B_{0}^{2}$, $B_{0}^{4}$, and $B_{4}^{4}$, we combine the effects of these crystal field parameters into a single parameter $\Delta$ for simplicity. 
The effect of the magnetic field is represented by the Zeeman coupling in $\mathcal{H}_{\rm Z}$, where we renormalize the Land\'e $g$ factor into $h$ for simplicity. 

In the following calculations, we set $\alpha=0.38$, which leads to the easy-axis anisotropic interaction in $\mathcal{H}_{\text{ex}}$, since the easy-axis magnetic anisotropy tends to stabilize the SkL in centrosymmetric magnets~\cite{leonov2015multiply, Hayami_PhysRevB.99.094420, hayami_PhysRevB.105.174437}.
Then, we construct the magnetic phase diagram by varying $\Delta$ and $h$.

\section{METHOD} \label{sec:method}
In this section, we present the
numerical method based on the mean-field calculations in Sec.~\ref{sec: Mean-Field Method}.
Then, we present physical quantities to identify obtained magnetic phases in Sec.~\ref{sec: Physical quantities}.

\subsection{Mean-Field Calculations}
\label{sec: Mean-Field Method}
In order to search for the instability toward the S-SkL in the effective localized model, we adopt the mean-field approximation to the exchange interaction Hamiltonian $\mathcal{H}_{\rm ex}$, where the two-body interactions reduce to a single-body problem
as follows~\cite{weiss1907hypothese}:
\begin{align}\label{eq:H_ex_MF}
    \mathcal{H}^{\rm MF}_{\text{ex}
    } 
    = - \sum_{\langle i,j\rangle} \mathcal{J}_{ij}\,( 
        \bm{J}_i \cdot \langle \bm{J}_j \rangle +\bm{J}_j \cdot \langle \bm{J}_i \rangle -
       \langle \bm{J}_i \rangle\cdot \langle \bm{J}_j \rangle).
\end{align}
Then, one can directly diagonalize the total Hamiltonian and obtain the eigenvalues $\epsilon_{n}$ and eigenstates $\ket{n}$. 
The expectation values of the observables $\mathcal{O}$ at the temperature $T$ is given by~\cite{Weise2008}
\begin{align}\label{eq:exactDia}
    \langle \mathcal{O} \rangle 
    = \frac{1}{Z} \sum_{n} \exp(- \epsilon_{n}/T) 
      \bra{n}\mathcal{O}\ket{n},
\end{align}
where $Z$ is the partition function,
\begin{align}\label{eq:partitionZ}
    Z = \sum_{n} \exp(-\epsilon_{n}/T), 
\end{align}
where we set the Boltzmann constant $k_{\rm B}$ to unity. The unit of the temperature $T$ is the exchange interaction $J_1$.

We iterate the mean-field calculations until both the free energy and mean values of magnetic moments at each lattice site converge to the precision of $10^{-6}$.
Since we consider the situation where the ordering wave vectors are $\bm{Q}_1$, we set a $6\times 6 $ unit cell under the periodic boundary conditions along both $x$ and $y$ directions. 
When the S-SkL characterized by the superposition of $\bm{Q}_1$ and $\bm{Q}_2$ is realized, two magnetic unit 
cells are included in the $6\times 6$ unit cell.

In addition, we set the initial configurations of $\bm{J}_i$ as follows: 
We set up 
one ferromagnetic configuration, one single-$Q$ spiral configuration,
one double-$Q$ spiral configuration, 10 conical configurations with different polar angles, and 200 SkL configurations that follow the formula in Ref.~\cite{UtesovPhysRevB.103.064414} in addition to approximately 200 random configurations. 
In the presence of the magnetic field $h > 0$, we additionally
adopt 200 lowest-free-energy converged solutions from the previous calculation. Once the calculation for a given crystal field splitting $\Delta$ is completed, approximately 10 converged results around each $h$ are also used as initial states for the subsequent calculation at each $h$. 
Furthermore, we introduce fluctuations in each converged magnetic moment at every site along $x$-, $y-$, and $z$-direction, ranging from $-0.3$ to $0.3$, in the aforementioned 10 converged results around each $h$. 
These 10 fluctuated configurations are then used alongside the 10 original converged results for further calculations.

\subsection{Physical Quantities}\label{sec: Physical quantities}
The obtained magnetic phases are identified by the 
structure factor
and scalar chirality. 
The structure factor in terms of the magnetic moment $\bm{J}_i$ is given by 

\begin{align}\label{eq:SSF}
    J(\bm{q}) = \frac{1}{N} \sum_{ 
        i,j 
        }^{}  
    \left \langle \bm{J}_{i} \right \rangle \cdot \left \langle  \bm{J}_{j} \right \rangle  \exp \left [ i\,  \bm{q}  \cdot   \left (\bm{r}_{i}-\bm{r}_{j}\right)  \right ],
\end{align}
where $N=36$ is the total number of sites.

In addition, we also calculate the scalar chirality in terms of the magnetic moments, which is defined by the triple scalar product as~\cite{XGWenPhysRevB.39.11413, Yi_PhysRevB.80.054416}
\begin{align}\label{eq:ScalarSpinChirality}
    \chi_{i} = \frac{1}{2} \sum_{\delta,\delta'= \pm1} \delta \delta' \left \langle \bm{J}_{i} \right \rangle \cdot \left(  \left \langle \bm{J}_{i+\delta\hat{x}} \right \rangle \times  \left \langle \bm{J}_{i+\delta'\hat{y}} \right \rangle \right),
\end{align}
where $\hat{x}$ ($\hat{y}$) represents a translation by the lattice constant along the $x$ ($y$) direction.
Summing the local scalar 
chirality over all sites on the lattice 
gives the net scalar 
chirality $\left\langle \chi \right\rangle$, defined as
\begin{align}\label{eq:meanChirality}
    \left \langle \chi \right \rangle = \frac{1}{N}\sum_{i}^{}\chi_{i}.
\end{align}

The vortex-like structure of a skyrmion is also characterized by the topological skyrmion number $N_{\text{sk}}$, which is defined via an integral of the solid angle~\cite{Rajaraman,braun2012topological}.
On a discrete lattice, this integral is replaced by a summation of local skyrmion densities, leading to the total skyrmion number as 
\begin{align}\label{eq:skyrmion number}
    N_{\text{sk}} = \frac{1}{2}\frac{1}{4\pi} \sum_{i} \Omega_{i}.
\end{align}
A prefactor $\tfrac{1}{2}$
is due to the two magnetic unit cells in the $6\times 6$ unit cell when the double-$Q$ magnetic-moment configurations emerge. Here,
$\Omega_{i} \in [-2\pi,\, 2\pi)$ is the skyrmion density~\cite{berg1981definition}, which is given by
\begin{align}\label{eq:S-SkLDensity}
   \Omega_{
   i} =
   \sum_{\delta,\delta'= \pm1} \arctan{\left( \frac{2\delta \delta' \left \langle \bm{j}_{
   i} \right \rangle \cdot \left( \left \langle \bm{j}_{
   j} \right \rangle \times \left \langle \bm{j}_{
   k} \right \rangle \right)}{\left( \left \langle \bm{j}_{
   i} \right \rangle + \left \langle \bm{j}_{
   j} \right \rangle + \left \langle \bm{j}_{
   k} \right \rangle \right)^{2} - 1} \right)}.
\end{align}
We normalize the magnetic moments as $\left \langle \bm{j}_{
i} \right \rangle  = \left \langle \bm{J}_{i} \right \rangle/|\left \langle \bm{J}_{i} \right \rangle|$; $j=i+\delta\hat{x}$ and $k=i+\delta'\hat{y}$.
Since the topological skyrmion number $N_{\text{sk}}$ counts how many times the
magnetic-moment configuration wraps around the unit sphere,
it becomes an integer. 
For example,
$\left|N_{\text{sk}}\right|=1$
indicates a single skyrmion in the magnetic unit cell.

Furthermore, we calculate the net magnetization, which is defined as the vector sum of all mean magnetic moments divided by $N$:
\begin{align}
    \left \langle \bm{J} \right \rangle = \frac{1}{N} \sum_{i}  \left \langle \bm{J}_i \right \rangle.
\end{align}
To quantitatively analyze the magnitude of the magnetization, we take the norm of $\left \langle \bm{J} \right \rangle$, denoted as $\left | \left \langle \bm{J} \right \rangle \right |$.

\section{RESULTS} \label{sec:results}
\begin{figure}[htbp]
    \begin{center}
    \includegraphics[width=8.5cm]{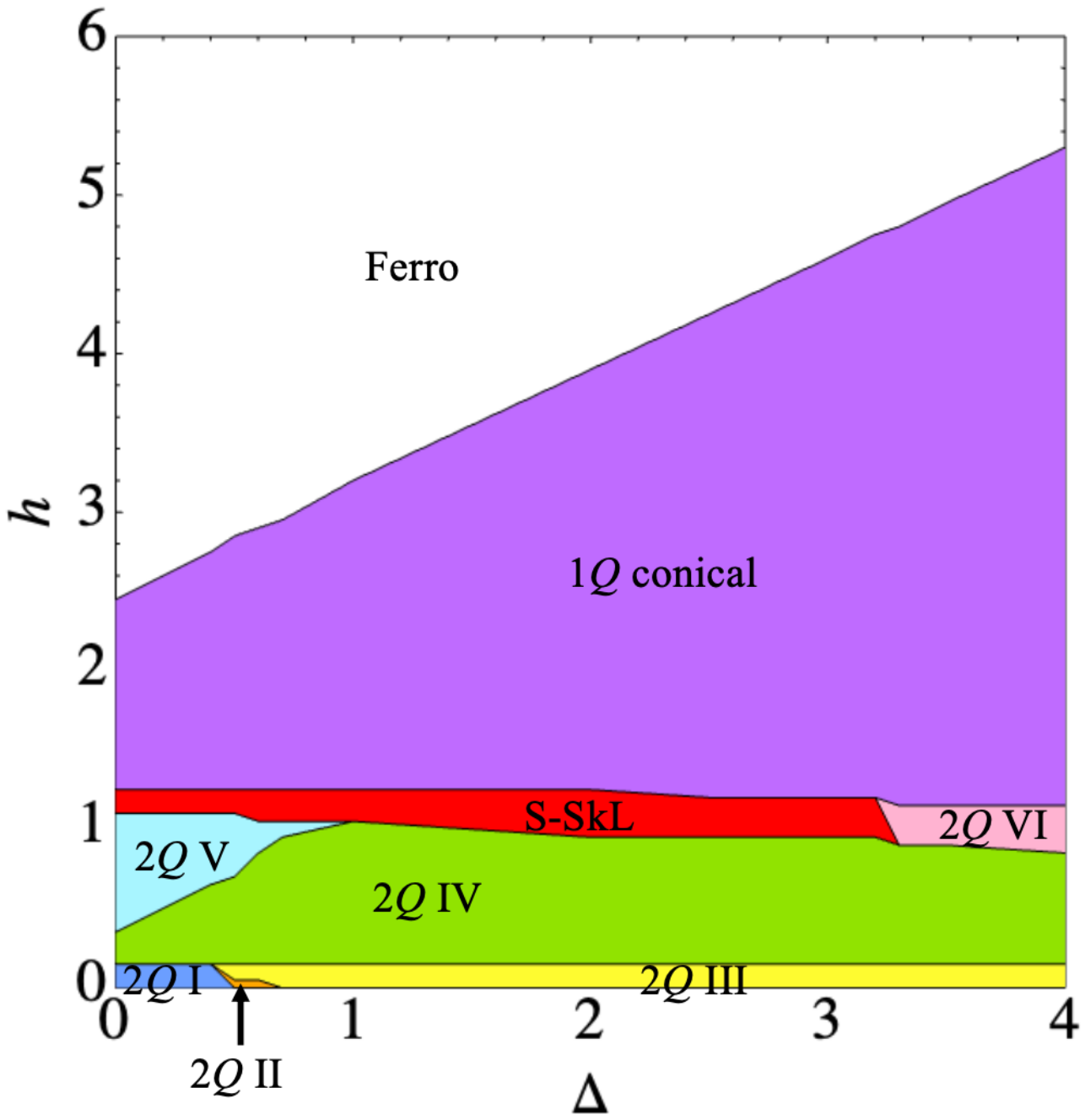}
    \caption{
    Phase diagram at low temperature ($T=0.05$) as functions of the magnetic field $h$ (vertical axis) and the
    crystal field splitting $\Delta$ between the two Kramers
    doublets (horizontal axis). Each colored region corresponds to a different low-temperature magnetic-moment configuration, including the 1$Q$ conical magnetic phase, various 2$Q$ phases (labeled I--VI), the S-SkL phase, 
    and the
    ferromagnetic phase (Ferro).
    }
    \label{fig_hDeltaPhaseDiagram}
    \end{center}
\end{figure}

\Figure{fig_hDeltaPhaseDiagram} shows the
$\Delta$--$h$ phase diagram
at low temperature 
$T=0.05$, which is obtained by the mean-field calculations.
The diagrams are
constructed by varying $h$ and $\Delta$, and independently performing the
iterations until the solutions converge for each initial magnetic-moment configuration.

\begin{figure*}[t!]
    \begin{center}
    \includegraphics[width=0.75 \hsize]{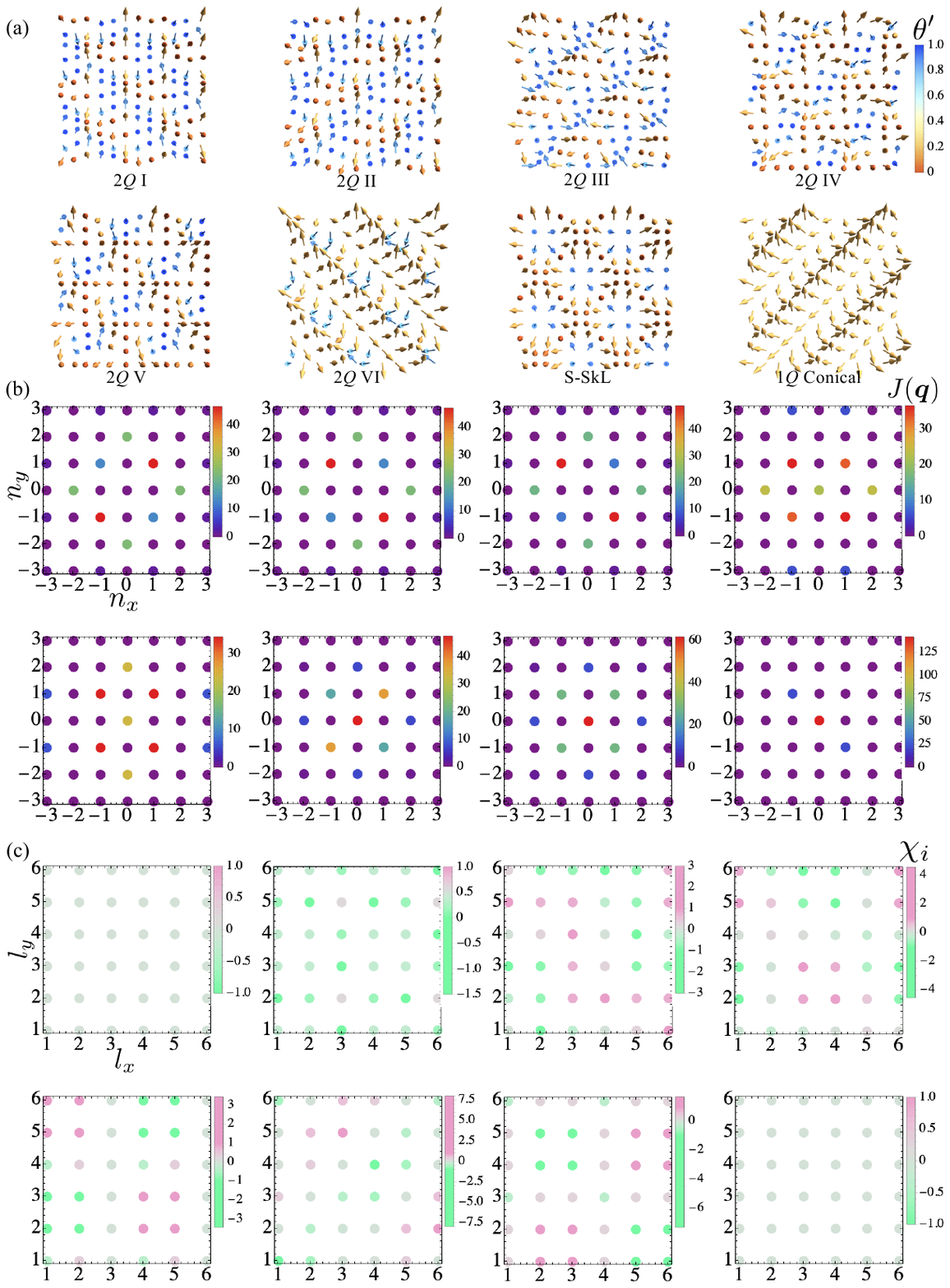}
    \caption{
    Panels (a)--(c) display three aspects of the real-space and momentum-space physical quantities:
    (a)
    Three-dimensional magnetic-moment configurations in a $6 \times 6$ unit cell, with magnetic moments drawn at each site and normalized in length for clarity. 
    The color scale encodes the normalized polar angle $\theta' = \theta / \pi$, varying continuously from 0 (north pole) to 1 (south pole).
    (b)
    Structure factor
    distributions $
    J(\mathbf{q})$ regarding the magnetic moments in
    momentum space. 
    The axis labels $n_{x}$ and $n_{y}$ denote the multiples of ${2\pi}/6$ with $-3 \leq n_{x}, n_{y} \leq 3$ in the first Brillouin zone.
    (c)
    Local scalar
    chirality $\chi_{i}$
    in the $6 \times 6$
    unit cell.
    }
    \label{fig_RealspaceSfChirality}
    \end{center}
\end{figure*}

We present eight types of real-space
magnetic-moment configurations for
each magnetic phase
in \Fig{fig_RealspaceSfChirality}(a). 
Although we perform the calculation
in the
$6 \times 6$
unit cell, we plot the $12 \times 12$ configurations by copying the original data for better visibility.
To further enhance the readability of the image, the 
magnetic-moment lengths $\left |\left \langle \bm{J}_{i} \right \rangle \right |$ are normalized only when plotting the
three-dimensional 
magnetic-moment configurations
\footnote{At $\Delta = 0$, the magnetic moment length $\left |\left \langle \bm{J}_{i} \right \rangle \right |$ varies from $2.47$ (2$Q$ I) to $2.50$ (Ferro) as the magnetic field increases. 
At $\Delta = 4$, $\left |\left \langle \bm{J}_{i} \right \rangle \right |$ varies from $2.27$ (2$Q$ III) to $2.49$ (Ferro) as the magnetic field increases.
}.
Additionally, we use the variation of the normalized polar angle $\theta'$
to depict the magnetic-moment orientation along the $z$-direction. The normalized polar angle $\theta'$ is defined as the ratio of the polar angle $\theta$ to $\pi$, with values ranging from 0 (north pole) to 1 (south pole) on the unit sphere.

\begin{figure*}[t!]
    \begin{center}
    \includegraphics[width=1 \hsize]{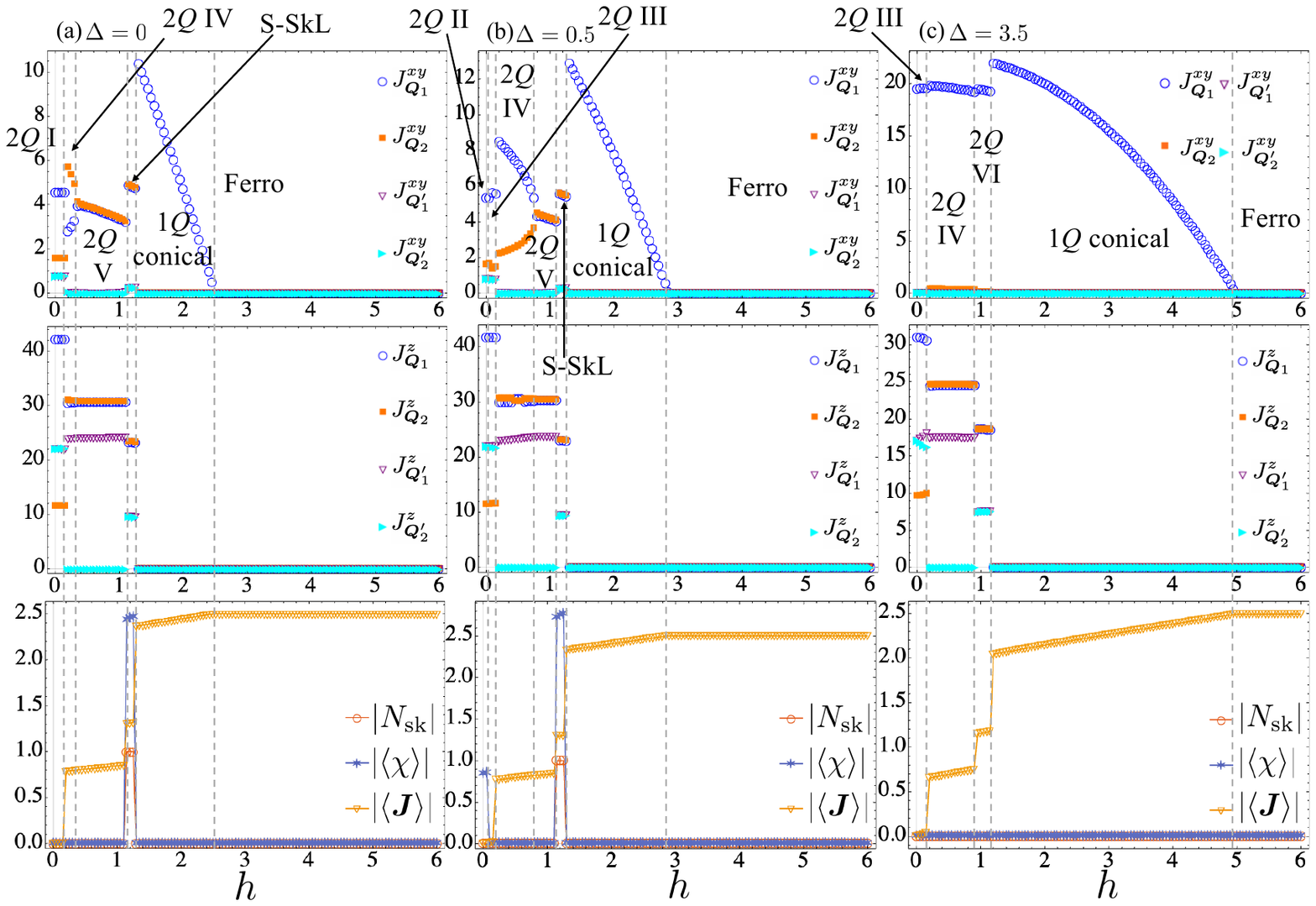}
    \caption{Panels (a)--(c) display three representative values of crystal field splitting i.e., $\Delta = 0, 0.5, 3.5$, respectively.
    The top and middle panels of (a)--(c) illustrate the evolution of the in-plane structure factor component $J^{xy}_{\bm{q}}$ and the out-of-plane component $J^{z}_{\bm{q}}$ with respect to the external magnetic field $h$ for the three representative crystal field splitting values.
    The bottom panels of (a)--(c) depict three physical quantities: the absolute topological skyrmion number $\left| N_{\text{sk}} \right|$, the absolute scalar chirality $\left|\langle \chi \rangle \right|$, and the magnetization magnitude $\left|\langle \bm{J} \rangle \right|$. These graphs reveal the interplay between crystal field effects and external magnetic field on the magnetic-moment configurations.}
    \label{fig_jxyjzChiraNskJ}
    \end{center}
\end{figure*}

Figure~\ref{fig_RealspaceSfChirality}(b) displays the
structure factor maps with respect to the magnetic moment in
momentum space, with a rainbow color scheme representing the intensity of each
structure factor.
The axis labels $n_{x}$ and $n_{y}$ denote the multiples of ${2\pi}/6$,
i.e., ${n_{x}2\pi}/6$ and ${n_{y}2\pi}/6$ for $-3 \leq n_x, n_y \leq 3$, in the first Brillouin zone.
To further classify the 
magnetic-moment
configurations, we decompose the structure factor into the in-plane one
$J^{xy}_{\bm{q}} =
J^{x}(\bm{q}) + 
J^{y}(\bm{q})$, and the out-of-plane
one, 
$J^{z}_{\bm{q}} =J^{z}(\bm{q})$.
These components are shown in \Fig{fig_jxyjzChiraNskJ}, under three representative values of crystal field splitting i.e., $\Delta = 0, 0.5, 3.5$, with the top and middle panels corresponding to \Fig{fig_jxyjzChiraNskJ}(a)--(c), respectively. 
Additionally, the bottom panels of \Fig{fig_jxyjzChiraNskJ}(a)--(c) presents three graphs depicting the magnetization magnitude $\left | \left \langle \bm{J} \right \rangle \right |$, the absolute net scalar chirality $\left |\left \langle \chi \right \rangle \right |$, and the absolute skyrmion number $\left | N_{\text{sk}} \right |$.
We 
analyze data in the same manner as in \Fig{fig_jxyjzChiraNskJ} and
list nonzero $J^{xy}_{\bm{q}}$ and $J^{z}_{\bm{q}}$ for the ordering wave vectors $\bm{Q}_1$ and $\bm{Q}_2$ as well as the higher-harmonic wave vectors $\bm{Q}'_1=\bm{Q}_1+\bm{Q}_2$ and $\bm{Q}'_2=\bm{Q}_1-\bm{Q}_2$ in each magnetic phase in Table~\ref{table:ssf}. 
In the double-$Q$ 
configuration with $\bm{Q}_1$ and $\bm{Q}_2$, the contributions from high-harmonic wave vectors $\bm{Q}'_1$ and $\bm{Q}'_2$ become nonzero owing to the superposition of $\bm{Q}_1$ and $\bm{Q}_2$. 

\begin{table}[htb!]
\centering
\caption{
Nonzero components of $
J_{\bm{Q}_\eta}$ and $
J_{\bm{Q}'_\eta}$ ($\eta=1,2$) in each phase. 
\label{table:ssf}}
\vspace{2mm}
\renewcommand{\arraystretch}{1.2}
\begin{tabular}{lccccccccccccccccccc}
\hline
\hline
phase &$J_{\bm{Q}_1}$, $J_{\bm{Q}_2}$ ($\bm{Q}_\eta \parallel [110]$) &$J_{\bm{Q}'_1}$, $J_{\bm{Q}'_2}$ ($\bm{Q}'_\eta \parallel [100]$) \\ \hline
1$Q$ conical & $J^{xy}_{\bm{Q}_1}$ & --\\  
\hline
S-SkL & $J^{xy}_{\bm{Q}_1}=J^{xy}_{\bm{Q}_2}$, $J^{z}_{\bm{Q}_1}=J^{z}_{\bm{Q}_2}$ & $J^{xy}_{\bm{Q}'_1}=J^{xy}_{\bm{Q}'_2}$, $J^{z}_{\bm{Q}'_1}=J^{z}_{\bm{Q}'_2}$ \\ 
\hline
2$Q$ I & $J^{xy}_{\bm{Q}_1}, J^{xy}_{\bm{Q}_2}$, $J^{z}_{\bm{Q}_1}, J^{z}_{\bm{Q}_2}$ &$J^{xy}_{\bm{Q}'_1}=J^{xy}_{\bm{Q}'_2}$, $J^{z}_{\bm{Q}'_1}=J^{z}_{\bm{Q}'_2}$ \\ 
2$Q$ II & $J^{xy}_{\bm{Q}_1}, J^{xy}_{\bm{Q}_2}$, $J^{z}_{\bm{Q}_1}, J^{z}_{\bm{Q}_2}$ & $J^{xy}_{\bm{Q}'_1}=J^{xy}_{\bm{Q}'_2}$, $J^{z}_{\bm{Q}'_1}=J^{z}_{\bm{Q}'_2}$ \\ 
2$Q$ III & $J^{xy}_{\bm{Q}_1}, J^{xy}_{\bm{Q}_2}$, $J^{z}_{\bm{Q}_1}, J^{z}_{\bm{Q}_2}$ & $J^{xy}_{\bm{Q}'_1}=J^{xy}_{\bm{Q}'_2}$, $J^{z}_{\bm{Q}'_1},J^{z}_{\bm{Q}'_2}$ \\ 
2$Q$ IV & $J^{xy}_{\bm{Q}_1},J^{xy}_{\bm{Q}_2}$, $J^{z}_{\bm{Q}_1}=J^{z}_{\bm{Q}_2}$ & $J^{z}_{\bm{Q}'_1}$ \\ 
2$Q$ V & $J^{xy}_{\bm{Q}_1}=J^{xy}_{\bm{Q}_2}$, $J^{z}_{\bm{Q}_1}=J^{z}_{\bm{Q}_2}$ & $J^{z}_{\bm{Q}'_1}$ 
\\ 
2$Q$ VI & $J^{xy}_{\bm{Q}_1}$, $J^{z}_{\bm{Q}_1}=J^{z}_{\bm{Q}_2}$ & $J^{z}_{\bm{Q}'_1}=J^{z}_{\bm{Q}'_2}$ 
\\ 
\hline\hline
\end{tabular}
\end{table}

In \Fig{fig_RealspaceSfChirality}(c), we show the local scalar
chirality maps across all the lattice sites in the $6\times 6$ unit cell, with a mint-colored scheme representing
its intensity
$\chi_{i}$ at each site. 
The axis labels $l_{x}$ and $l_{y}$ represent the lattice site indices with $1\leq l_x, l_y \leq 6$.
From the data of the structure factor in \Fig{fig_RealspaceSfChirality}(b), \Fig{fig_jxyjzChiraNskJ}, and Table~\ref{table:ssf} and the scalar chirality in \Fig{fig_RealspaceSfChirality}(c) and 
\Fig{fig_jxyjzChiraNskJ}, we distinguish magnetic phases in the phase diagram.

Among the obtained phases, the most important observation is the appearance of the S-SkL in the intermediate-field region of the phase diagram in Fig.~\ref{fig_hDeltaPhaseDiagram}.
Specifically,
this phase is stabilized when the
crystal field splitting between the two doublets is approximately
$0 \leq \Delta < 3.5$, and the magnetic field is around $h \simeq 1.1$. 
The region of the S-SkL becomes the largest when $\Delta \simeq 2$.
This stability tendency indicates that the emergence of this S-SkL is owing to the multi-orbital effect that is often neglected by the classical spin model in previous studies. 
Indeed, the S-SkL disappears for large $\Delta$, where the multi-orbital effect is negligible. 
Such a tendency is consistent with previous studies in the classical spin models; the S-SkL has been found only in the presence of the compass-type magnetic anisotropy~\cite{wang_PhysRevB.103.104408} or dipolar interaction~\cite{UtesovPhysRevB.103.064414} or further neighbor interactions beyond the thrid-neighbor spins~\cite{Okigami_PhysRevB.110.L220405}. 

The behaviors of the magnetic moments in real and momentum spaces in the obtained S-SkL are similar to those in the classical spin models~\cite{hayami_PhysRevB.105.174437}.
The S-SkL exhibits a
net component of the scalar 
chirality, $\left\langle \chi \right\rangle \simeq 3$
, which is influenced by the alterable
magnetic-moment
length and thus varies as $\Delta$ changes. 
The absolute value of the topological skyrmion number is $\left|N_{\text{sk}}\right| = 1$, where the energy of the 
magnetic-moment
configuration with $N_{\text{sk}} = 1$ (antiskyrmion) is degenerated with that with $N_{\text{sk}} = -1$ (
skyrmion) owing to the absence of the bond-dependent anisotropy~\cite{Hayami_doi:10.7566/JPSJ.89.103702}; it is noted that the helicity of skyrmion is not fixed, which is also attributed to the absence of the bond-dependent anisotropy.
The structure factor exhibits the fourfold-symmetric peak structures as follows:
$J^{xy}_{\bm{Q}_1} = J^{xy}_{\bm{Q}_2}$, $J^{z}_{\bm{Q}_1} = J^{z}_{\bm{Q}_2}$, $J^{xy}_{\bm{Q}'_1} = J^{xy}_{\bm{Q}'_2}$, and $J^{z}_{\bm{Q}'_1} = J^{z}_{\bm{Q}'_2}$ as shown in \Fig{fig_RealspaceSfChirality}(b) and \Fig{fig_jxyjzChiraNskJ}. 

\begin{figure}[htbp]
    \begin{center}
    \includegraphics[width=8.5cm]{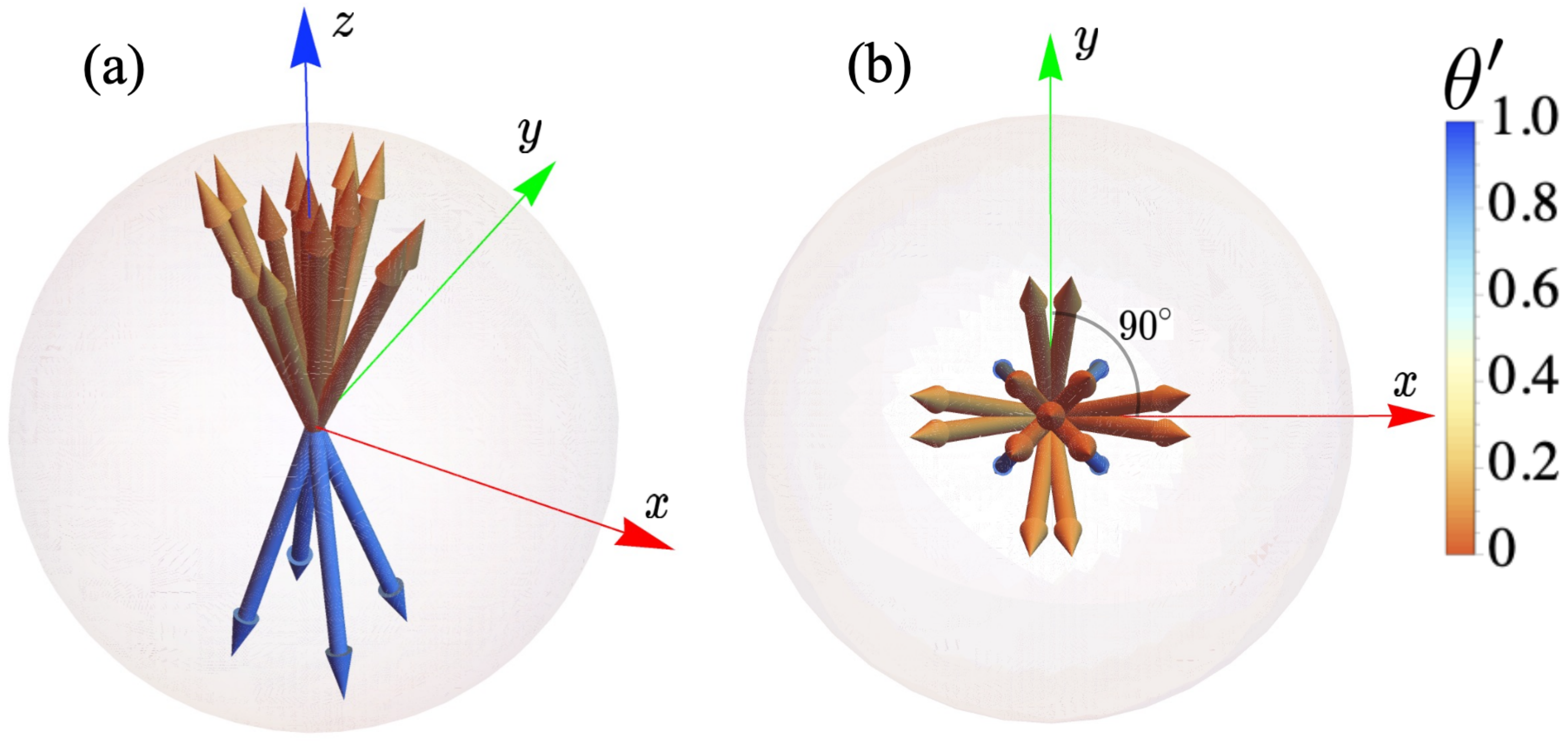}
    \caption{A schematic of real-space magnetic moments plotted on the surface of a unit sphere, where the magnetic moments forming the S-SkL are distributed primarily near the north and south poles. (a): Default view. (b): Top view.
    }
    \label{fig_SkLonSphere}
    \end{center}
\end{figure}

The magnetic-moment configuration of the S-SkL is approximately expressed as  
\begin{align}
\label{eq:SkX}
\bm{J}_i \propto 
\left(
    \begin{array}{c}
   - \cos \mathcal{Q}_1 +  \cos \mathcal{Q}_2 \\
   - \cos \mathcal{Q}_1 -  \cos \mathcal{Q}_2 \\
   - a_{z} (\sin \mathcal{Q}_1+\sin \mathcal{Q}_2)+ \tilde{h}_z
          \end{array}
  \right)^{\rm{T}}, 
\end{align}
where $\mathcal{Q}_\eta = \bm{Q}_\eta \cdot \bm{r}_i + \theta_\eta$ for $\eta=1,2$.
The $a_z$ and $\tilde{h}_z$ stand for the parameters related to the degree of the easy-axis magnetic anisotropy and uniform magnetization, respectively.  
By empirically choosing the parameters to approximately reproduce the real-space magnetic-moment configuration of the S-SkL obtained by the mean-field calculations in \Fig{fig_RealspaceSfChirality}(a), we obtain 
$a_{z} \simeq 4$, $\tilde{h}_z\simeq 3.5$, $\theta_{1} \in (0,\tfrac{\pi}{4})$ [or $(-\tfrac{\pi}{4},0)$], and $\theta_{2} \in (-\tfrac{\pi}{4},0)$ [or $(0,\tfrac{\pi}{4})$], normalizing the magnetic moments and scaling them by the length $l \simeq 2.4$.
The relatively large value of $a_z$ indicates that the energy gain
from the easy-axis magnetic anisotropy plays an important role in stabilizing the S-SkL. 
Indeed, the magnetic moments consisting of the S-SkL are distributed in the vicinity of the north and south poles in the unit sphere, as shown in \Fig{fig_SkLonSphere}. The top view in \Fig{fig_SkLonSphere}(b) demonstrates the fourfold rotational symmetry of the S-SkL. 

\begin{figure}[htbp]
    \begin{center}
    \includegraphics[width=8.5cm]{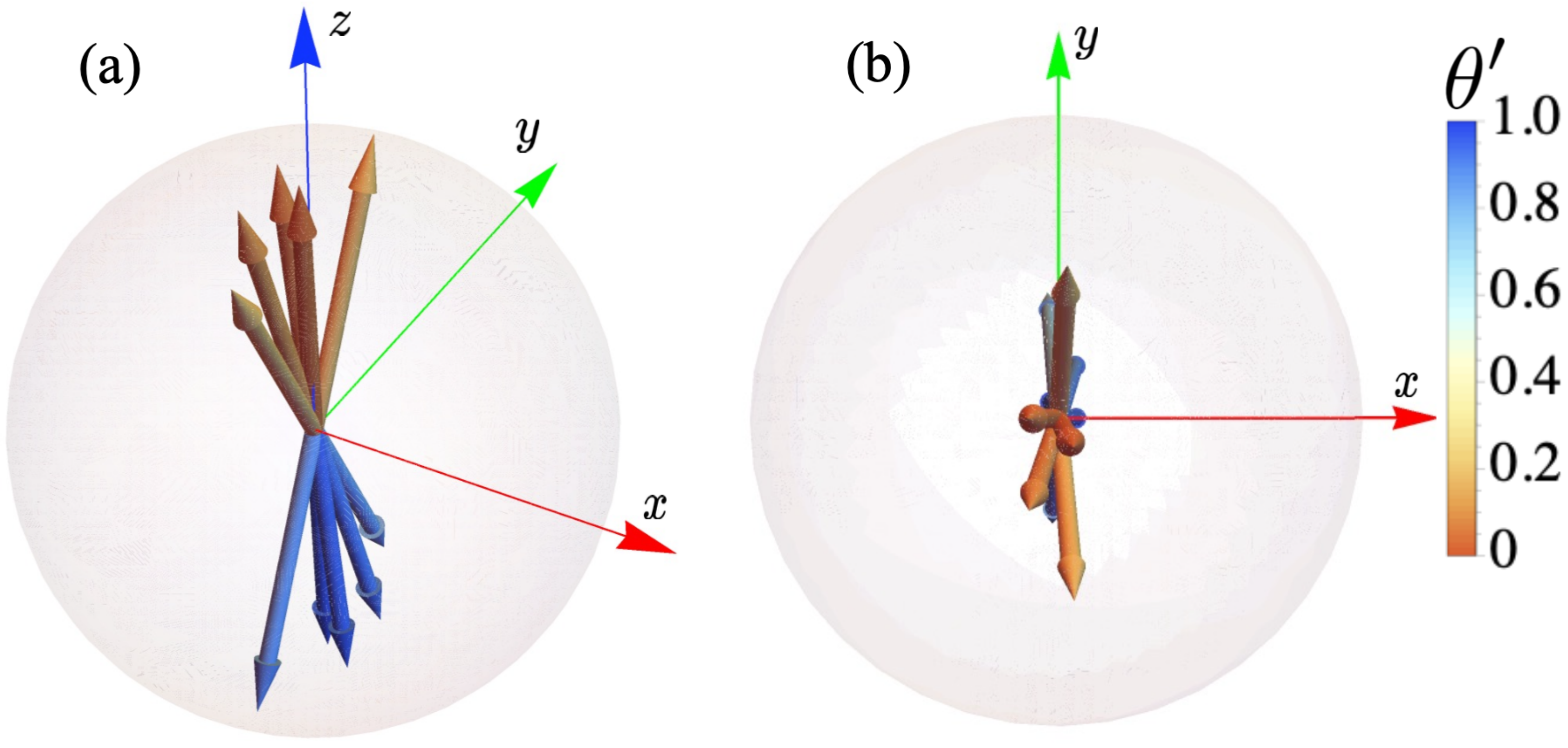}
    \caption{A schematic of real-space magnetic moments plotted on the surface of a unit sphere, illustrating the
    2$Q$ II state. The magnetic moments near the north and south poles of the unit sphere are nearly collinear. (a): Default view. (b): Top view. 
    }
    \label{fig_2QIIonSphere}
    \end{center}
\end{figure}

In addition to the S-SkL, we find another double-$Q$ state with a net scalar chirality, which is denoted as the 2$Q$ II state in the phase diagram in Fig.~\ref{fig_hDeltaPhaseDiagram}. 
This state is stabilized in the narrow regions of $\Delta$ and $h$: 
$0.5 \lesssim \Delta  \lesssim 0.64$
and 
$0 \leq h \lesssim 0.15$.
Similar to the S-SkL, this state accompanies a net scalar chirality, as shown in the bottom panel of Fig.~\ref{fig_jxyjzChiraNskJ}(b). 
On the other hand, the skyrmion number of this phase fluctuates depending on $\Delta$; the skyrmion number is zero for $0.5 \lesssim \Delta \lesssim 0.6$, whereas it is one for $0.6 \lesssim \Delta \lesssim 0.64$. 
The reason why the different skyrmion numbers might be attributed to the almost coplanar magnetic-moment
configuration in the 2$Q$ II state. 
As shown in the real-space magnetic-moment distribution in Figs.~\ref{fig_2QIIonSphere}(a) and \ref{fig_2QIIonSphere}(b), almost all of the magnetic moments are distributed in the same plane, although there is a slightly out-of-plane component. 
This indicates the 
finite scalar chirality with the large solid angle owing to the strong easy-axis anisotropy, which leads to the sign change
sensitive to the moment direction. 
In addition, the magnetic-moment configuration of this state breaks the fourfold rotational symmetry, as clearly found in Fig.~\ref{fig_2QIIonSphere}(b). 
It is noted that this state also emerges thanks to the multi-orbital effect, which has not been reported in the classical spin model, where only the zero-field SkL phase with the skyrmion number of two has been found in centrosymmetric hosts~\cite{wang_PhysRevB.103.104408, hayami2022multiple}. 

Finally, let us briefly discuss the characteristics of the other magnetic phases in the phase diagram.
Except for the S-SkL
and the 2$Q$ II state, we identify one single-$Q$ state and
five types of other double-$Q$ states.
We summarize the nonzero components of $J_{\bm{Q}_\eta}$ and $J_{\bm{Q}'_\eta}$ in Table~\ref{table:ssf}. 
The 1$Q$ conical state with the intensity at $\bm{Q}_1$ or $\bm{Q}_2$ is stabilized across all $\Delta$ in the high-field region; the spiral plane lies in the $xy$ plane. 
The
critical value of the magnetic field to this state
is approximately $h\simeq 1.3$. 
This state turns into the ferromagnetic state, where the critical magnetic field
varies linearly with $\Delta$, as shown in \Fig{fig_hDeltaPhaseDiagram}.
The 2$Q$ I state is stabilized at 
$0 \leq \Delta < 0.5$
and in the small magnetic field
around $0 \leq h < 0.2$, occupying a narrow region of the phase diagram. 
The 2$Q$ I state is the only double-$Q$ state that lacks local scalar chirality $\chi_i$, as shown in \Fig{fig_RealspaceSfChirality}(c). 
In other words, this state is characterized by the coplanar magnetic-moment configuration, as shown by the real-space configuration in Fig.~\ref{fig_RealspaceSfChirality}(a).
As the crystal field splitting $\Delta$ increases,
the 2$Q$ I state within $h < 0.2$ undergoes continuous deformation
through the 2$Q$ II state with a net scalar chirality, and eventually transform into the 2$Q$ III state.
The 2$Q$ III state is stabilized
for $\Delta > 0.64$. 
The in-plane and out-of-plane components of the structure factor in the 2$Q$ III state closely resemble those of the 2$Q$ I and 2$Q$ II states, as shown in \Fig{fig_RealspaceSfChirality}(b) and \Fig{fig_jxyjzChiraNskJ}, although the relation of $J^{z}_{\bm{Q}'_1}=J^{z}_{\bm{Q}'_2}$ in the 2$Q$ I and 2$Q$ II states no longer holds. 
When $\Delta$ increases, the values of $J^{z}_{\bm{Q}'_1}$, $J^{xy}_{\bm{Q}'_2}$, and $J^{xy}_{\bm{Q}_2}$ rapidly decrease
, as shown in the top panel of \Fig{fig_jxyjzChiraNskJ}(c).
The 2$Q$ IV state is stabilized for all $\Delta$
up to $\Delta = 4$, emerging at $h > 0.15$. 
In the region $0 \leq \Delta \lesssim 1$, the critical magnetic field separating the 2$Q$ IV and 2$Q$ V states increases linearly until $h \simeq 0.95$. 
Beyond this point, the 2$Q$ V state disappears, and the upper bound of the 2$Q$ IV state remains at $h \simeq 1.0$, as shown in \Fig{fig_hDeltaPhaseDiagram}. 
The lack of fourfold rotational symmetry in the 2$Q$ IV state is due to $J_{\bm{Q}_{1}}\ne J_{\bm{Q}_{2}}$, as demonstrated in \Fig{fig_RealspaceSfChirality}(b) and \Fig{fig_jxyjzChiraNskJ}.
The 2$Q$ V state is stabilized at $0 \leq \Delta \lesssim 1$. 
As $\Delta$ increases, the region occupied by the 2$Q$ V state shrinks. 
Although this state satisfies $J_{\bm{Q}_{1}} = J_{\bm{Q}_{2}}$, inhomogeneous contributions from higher-harmonic wave vectors break the fourfold rotational symmetry, as seen in \Fig{fig_RealspaceSfChirality}(b) and \Fig{fig_jxyjzChiraNskJ}(a)--(b).
The 2$Q$ VI state emerges for $3.2 < \Delta$ within the magnetic field ranged around $0.9 \lesssim h \lesssim 1.15$.
This state replaces the S-SkL region in the phase diagram. 
As shown in \Fig{fig_RealspaceSfChirality}(b) and \Fig{fig_jxyjzChiraNskJ}(c), high-harmonic wave vectors contribute to the fourfold rotational symmetry, as indicated by the relations $J^{z}_{\bm{Q}'_1} = J^{z}_{\bm{Q}'_2}$. 
However, the inequality $J^{xy}_{\bm{Q}_1} \ne J^{xy}_{\bm{Q}_2}$ signifies the breaking of this symmetry.
In this way, the competing interactions in
multi-orbital systems give rise to a rich variety of multi-$Q$ states as the lowest-energy configurations.

\section{SUMMARY} \label{sec:discussion and summary}
To summarize,
we have
investigated the emergence of
the S-SkL on a centrosymmetric
square lattice
by employing mean-field calculations with an emphasis on the multi-orbital degree of freedom.
By
taking into account the effects of the atomic spin--orbit coupling and
tetragonal crystalline electric field, we have constructed an effective localized
model consisting of two Kramers doublets with
easy-axis magnetic anisotropy.
Then, we have clarified the low-temperature phase diagram, which includes 
the S-SkL with the skyrmion number of one and 
the double-$Q$ state with the nonzero scalar chirality (2$Q$ II).
We have shown that the multi-orbital effect assists the stabilization of
the S-SkL and the 2$Q$ II state by changing the crystal-field splitting between two Kramers doublets.
We also found that a variety of double-$Q$ states can be realized in the multi-orbital model. 
Our study reveals a possibility of stabilizing S-SkLs in $4f$-electron systems with
a finite orbital angular momentum $\bm{L} \neq \bm{0}$, as found in
the Ce$^{3+}$ ion 
on a centrosymmetric
square lattice. This finding opens a new avenue for subsequent studies on skyrmion-hosting materials 
with the orbital degree of freedom.

\begin{acknowledgments}

Y. Zha would like to express his gratitude to Y. Ogawa, T. Shirato, and T. Yamanaka from Hokkaido University for their fruitful discussions. Y. Zha is also grateful to P.L. Lu and Y.J. Peng from Fudan University for their valuable comments on an early version of this paper. 
This research was supported by JSPS KAKENHI Grants Numbers JP21H01037, JP22H00101, JP22H01183, JP23H04869, JP23K03288, JP23K20827, and by JST CREST (JPMJCR23O4) and JST FOREST (JPMJFR2366).

\end{acknowledgments}

\appendix
\section{Derivation of the atomic bases}
\label{app1}

In this Appendix, we show the derivation of the atomic bases in Eq.~(\ref{eq:2KramersDoublets}) in the main text. 
In a $4f$-electron wave function with the $f^1$ configuration to have the orbital angular momentum $L=3$ in the presence of the spherical symmetry, there are $(2J+1)$-fold degenerate states. 
Such a degeneracy is lifted when the relativistic spin--orbit coupling and the crystalline electric field are considered. 
First, we consider the effect of the spin--orbit coupling, whose Hamiltonian is given by 
\begin{align}\label{eq:LS}
    \mathcal{H}_{LS} = \lambda \,\bm{L}\cdot \bm{S},
\end{align}
where $\bm{L} = (L^{x}, L^{y}, L^{z})$ and $\bm{S} = (S^{x}, S^{y}, S^{z})$ stand for the orbital and spin angular momentum operators, respectively.
$\lambda$ represents the spin--orbit coupling constant. 
This leads to the energy splittings into the eightfold-degenerated multiplet with $J=\tfrac{7}{2}$ and sixfold-degenerated multiplet with $J= \tfrac{5}{2}$.
Hereafter, we focus on the $J= \tfrac{5}{2}$ multiplet by supposing the larger spin--orbit coupling. 

Next, we take into account the effect of the tetragonal crystalline electric field, which further splits the above-degenerated bands into three Kramers doublets. 
The crystal field Hamiltonian is given by 
\begin{align}\label{eq:Hcry}
\mathcal{H}_{\text{cry}} = B_{2}^{0}O_{2}^{0} + B_{4}^{0}O_{4}^{0} + B_{4}^{4}O_{4}^{4}, 
\end{align}
where $B_{2}^{0}$, $B_{4}^{0}$, and $B_{4}^{4}$ are the crystal field parameters. 
The Stevens operator, $O_{n}^{m}$, is defined by~\cite{stevens1952matrix,hutchings1964point}
\begin{eqnarray}
\label{eq:stevens}
O_{2}^{0} &=& 3J_{z}^{2} - \bm{J}^{2},  \nonumber\\
O_{4}^{0} &=& 35J_{z}^{4} - 30\bm{J}^{2}J_{z}^{2} + 25J_{z}^{2} - 6\bm{J}^{2} + 3\bm{J}^{4}, \\
O_{4}^{4} &=& \frac{1}{2} \left[ (J^{+})^4 + (J^{-})^4 \right]. \nonumber
\end{eqnarray}

By applying $\mathcal{H}_{\text{cry}}$ to the sixfold $J= \tfrac{5}{2}$ basis, the matrix element of the crystal field Hamiltonian is expressed as 
\begin{widetext}
\begin{align}
\label{eq:h_cry}
\left(
\begin{array}{cccccc}
 10 (B_{2}^{0}+6 B_{4}^{0}) & 0 & 0 & 0 & 12 \sqrt{5} B_{4}^{4} & 0 \\
 0 & -2 (B_{2}^{0}+90 B_{4}^{0}) & 0 & 0 & 0 & 12 \sqrt{5} B_{4}^{4} \\
 0 & 0 & -8 (B_{2}^{0}-15 B_{4}^{0}) & 0 & 0 & 0 \\
 0 & 0 & 0 & -8 (B_{2}^{0}-15 B_{4}^{0}) & 0 & 0 \\
 12 \sqrt{5} B_{4}^{4} & 0 & 0 & 0 & -2 (B_{2}^{0}+90 B_{4}^{0}) & 0 \\
 0 & 12 \sqrt{5} B_{4}^{4} & 0 & 0 & 0 & 10 (B_{2}^{0}+6 B_{4}^{0}) \\
\end{array}
\right),
\end{align}
\end{widetext}
where the order of the basis wave function is represented by $\ket{5/2,J^{z}}$, from $\ket{5/2,5/2}$ down to $\ket{5/2,-5/2}$.
The eigenvalues are given by 
\begin{align}
    &-8 (B_{2}^{0}-15 B_{4}^{0}), \nonumber \\
    &- 6 \sqrt{(B_{2}^{0}+20 B_{4}^{0})^2 + 20 (B_{4}^{4})^2} +4 B_{2}^{0}  - 60 B_{4}^{0}, \nonumber \\
    &6 \sqrt{(B_{2}^{0}+20 B_{4}^{0})^2 + 20 (B_{4}^{4})^2} + 4 B_{2}^{0} - 60 B_{4}^{0}, \nonumber 
\end{align}
where the corresponding eigenstates are given by
\begin{align}
    \label{eq:3KramersDoublets}
    \begin{cases}
        \ket{\Gamma_{t6\pm}} 
        = \ket{\tfrac{5}{2},\pm \tfrac{1}{2}},
        \\[10pt]
        \ket{\Gamma_{t7\pm}^{(1)}} 
        = \alpha \ket{\tfrac{5}{2},\pm \tfrac{5}{2}} 
          - \beta \ket{\tfrac{5}{2},\mp \tfrac{3}{2}}, 
        \\[10pt]
        \ket{\Gamma_{t7\pm}^{(2)}} 
        = \beta \ket{\tfrac{5}{2},\pm \tfrac{5}{2}} 
          + \alpha \ket{\tfrac{5}{2},\mp \tfrac{3}{2}},
    \end{cases}
\end{align}
respectively. 
Here, $\alpha$ and $\beta$ are given in the main text.

In the localized
model, we appropriately choose the crystal field parameters so that two bases $\ket{\Gamma_{t7\pm}^{(1)}}$ and $\ket{\Gamma_{t7\pm}^{(2)}}$ are relevant in the targeting physical space for simplicity. 
We then set the atomic energy levels of $\ket{\Gamma_{t7\pm}^{(1)}}$ and $\ket{\Gamma_{t7\pm}^{(2)}}$ to 0 and $\Delta$, respectively.

\section{Other superposition coefficients $\alpha$}
\label{app2}

\begin{figure*}[t!]
    \begin{center}
    \includegraphics[width=0.89\hsize]{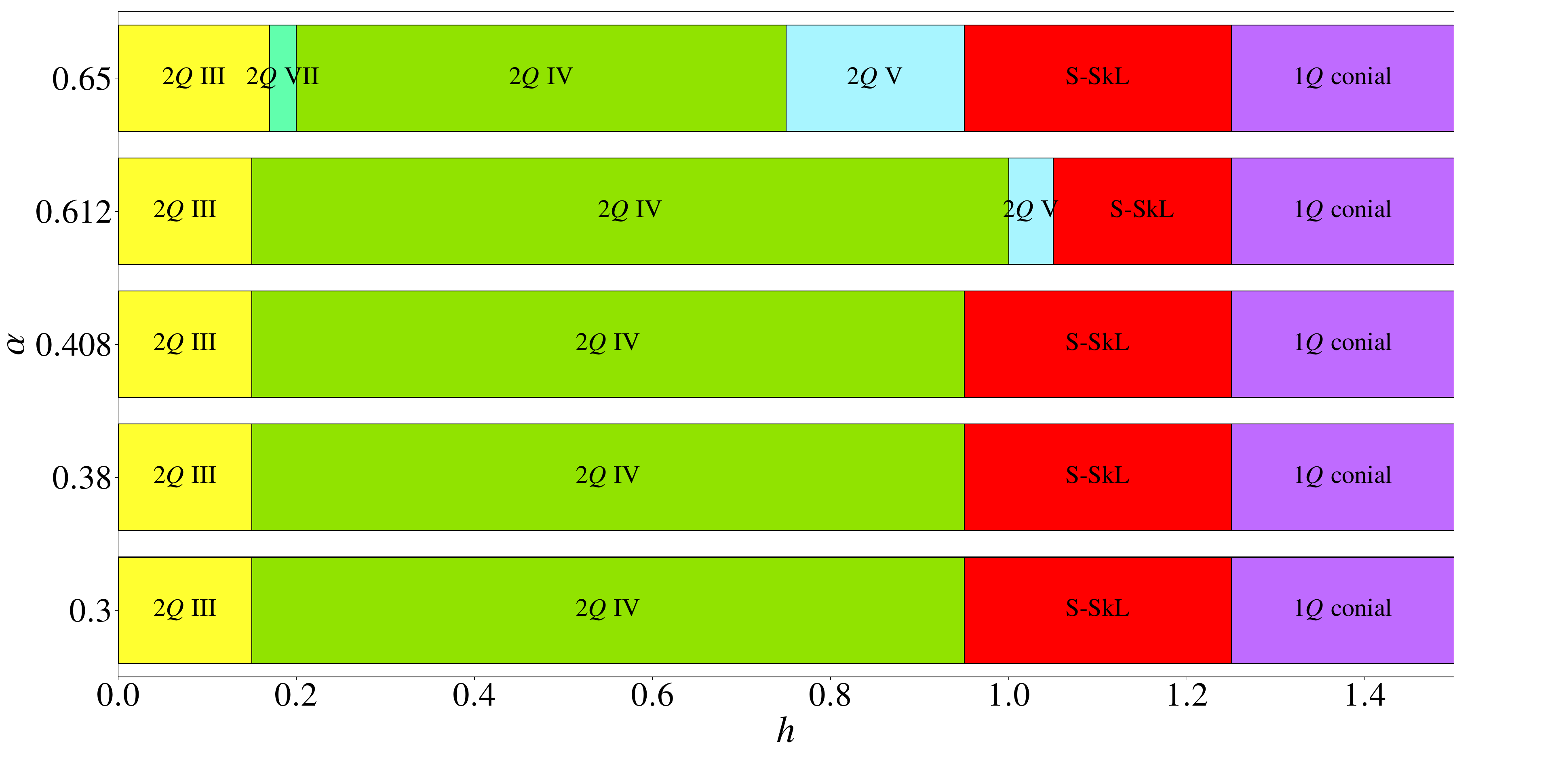}
    \caption{Phase diagrams for several representative values of
    $\alpha$ under
    $\Delta = 2$. Each panel corresponds to a different value of $\alpha$: $0.3$, $0.38$, $0.408$, $0.612$, and $0.65$ (from bottom to top). The vertical axis denotes $\alpha$, while the horizontal axis denotes the external magnetic field $h$, ranging from $0$ to $1.5$.}
    \label{fig_alphaDiagram}
    \end{center}
\end{figure*}

In this Appendix, we
discuss the stability of S-SkL for different choices of $\alpha$ in \Eq{eq:2KramersDoublets}  at $\Delta=2$.
Figure~\ref{fig_alphaDiagram} provides an overview of the phase diagrams for
$\alpha = 0.3, 0.38, 0.408, 0.6124,$ \text{and} $0.65$; the result at $\alpha=0.38$ is the same as that in Fig.~\ref{fig_hDeltaPhaseDiagram}.
In all the cases, the S-SkL phase appears in the
phase diagram. 
Especially, it is noteworthy that the S-SkL survives even when the $xy$-plane components in $\bm{J}$ are larger than the $z$-axis for $\alpha=0.6124$ and $0.65$. 
We find that the essence to stabilize the S-SkL for $\alpha=0.6124$ and $0.65$ is the off-diagonal component of $\bm{J}$ between the $\ket{\Gamma_{t7 \pm}^{(1)}}$ and $\ket{\Gamma_{t7 \pm}^{(2)}}$ levels, which favors the easy-axis-type magnetic-moment configuraiton.

We also
obtain an additional phase (2$Q$ VII)
for
$\alpha=0.65$
, which appears in the narrow field region around $0.17 \lesssim h \lesssim 0.2$. 
This phase is characterized by a nonzero net magnetization $|\langle \bm{J} \rangle| \simeq 0.21$ as well as the double-$Q$ modulations at $\bm{Q}_1$ and $\bm{Q}_2$, which varies with the applied magnetic field.
We list the nonzero components of $J_{\bm{Q}_\eta}$ and $J_{\bm{Q}'_\eta}$ ($\eta=1,2$) in the 2$Q$ VII phase in Table~\ref{table:ssf2QVII}.

\begin{table}[htb!]
\centering
\caption{
Nonzero components of $
J_{\bm{Q}_\eta}$ and $
J_{\bm{Q}'_\eta}$ ($\eta=1,2$) in phase 2$Q$ VII. 
\label{table:ssf2QVII}}
\vspace{2mm}
\renewcommand{\arraystretch}{1.2}
\begin{tabular}{lccccccccccccccccccc}
\hline
\hline
phase &$J_{\bm{Q}_1}$, $J_{\bm{Q}_2}$ ($\bm{Q}_\eta \parallel [110]$) &$J_{\bm{Q}'_1}$, $J_{\bm{Q}'_2}$ ($\bm{Q}'_\eta \parallel [100]$) \\ \hline
2$Q$ VII & $J^{xy}_{\bm{Q}_1} = J^{xy}_{\bm{Q}_2}$, $J^{z}_{\bm{Q}_1} = J^{z}_{\bm{Q}_2}$ &$J^{xy}_{\bm{Q}'_1}$, $J^{z}_{\bm{Q}'_1}, J^{z}_{\bm{Q}'_2}$ \\ 
\hline\hline
\end{tabular}
\end{table}

\bibliography{apssamp}

\end{document}